\documentclass[11pt]{elsart}
\usepackage[dvips]{graphicx}

\begin{document}
\begin{frontmatter}

\title{Relations between Average Distance, Heterogeneity and Network Synchronizability}

\author{Ming Zhao$^{a}$ }
\author{Tao Zhou$^{a,b}$ }
\ead{zhutou@ustc.edu}
\author{Bing-Hong Wang$^{a}$ }
\ead{bhwang@ustc.edu.cn}
\author{Gang Yan$^{b}$ }
\author{Hui-Jie Yang$^{a}$ }
\author{and Wen-Jie Bai$^{c}$ }

\address
{$^{a}$ Department of Modern Physics and Nonlinear Science Center,
University of Science and Technology of China,  Hefei Anhui,
230026, PR China }

\address
{$^{b}$ Department of Electronic Science and
Technology, University of Science and Technology of China,  Hefei
Anhui, 230026, PR China}

\address
{$^{c}$ Department of Chemistry, University of Science and
Technology of China,  Hefei Anhui, 230026, PR China}

\begin{abstract}
By using the random interchanging algorithm, we investigate the
relations between average distance, standard deviation of degree
distribution and synchronizability of complex networks. We find that
both increasing the average distance and magnifying the degree
deviation will make the network synchronize harder. Only the
combination of short average distance and small standard deviation
of degree distribution that ensures strong synchronizability. Some
previous studies assert that the maximal betweenness is a right
quantity to estimate network synchronizability: the larger the
maximal betweenness, the poorer the network synchronizability. Here
we address an interesting case, which strongly suggests that the
single quantity, maximal betweenness, may not give a comprehensive
description of network synchronizability.

\begin{keyword}
Synchronizability\sep Complex Networks \sep Average Distance\sep
Heterogeneity
\PACS 89.75.-k\sep 05.45.Xt
\end{keyword}
\end{abstract}

\date{}
\end{frontmatter}

\section{Introduction}
A variety of systems in nature can be described by complex
networks and the most important statistical features of complex
networks are the small-world effect and scale-free
property\cite{Review1,Review2,Review3,Review4}. Networks that have
small average distance as random networks and large clustering
coefficient as regular ones are called small-world
networks\cite{WS}. And the scale-free property means the degree
distribution of networks obeys the power-law form\cite{SFN}. One
of the ultimate goals of researches on complex networks is to
understand how the structure of complex networks affects the
dynamical process taking place on them, such as traffic
flow\cite{Traffic1,Traffic2,Traffic3,Traffic4,Traffic5}, epidemic
spread\cite{Epidemic1,Epidemic2,Epidemic3,Epidemic4,Epidemic5,Epidemic6},
cascading behavior\cite{Cascade1,Cascade2,Cascade3}, and so on.

The large networks of coupled dynamical systems that exhibit
synchronized state are subjects of great interest. Previous
studies have demonstrated that scale-free and small-world networks
are much easier to synchronize than regular
lattice\cite{Easier1,Easier2,Easier3,Easier4}. Then what makes
complex networks synchronize so easily? It is intuitively believed
that shorter average distance predicts better
synchronizability\cite{Easier2,Easier3,Zhou2005}. However, it is
found that to decrease average distance will make some complex
networks synchronize even harder\cite{Betweenness1}. More
bewilderingly, a very recent work suggests that on some
synchronization systems, the synchronizability is independent of
the average distance\cite{Hasegawa}. Some authors also addressed
that the homogeneous distribution of degree will lead to better
synchronizability. Hong \emph{et al.}\cite{Betweenness2}
investigate the relationship between network synchronizability and
various topological ingredients, including average distance,
heterogeneity, and betweenness of Watts-Strogatz (WS)
networks\cite{WS}. They suggest the maximal betweenness a right
indicator for synchronizability. This tentative conclusion has
been widely accepted now\cite{Cluster,Betweenness3,Oh}. Recently,
several researchers examine the effect of clustering coefficient
on the synchronization by using Kuramoto model\cite{Cluster} or
master stability function\cite{WuXiang} and find that increasing
clustering coefficient will hinder the global synchronization. All
the four topological ingredients, average distance, heterogeneity
(measured by the standard deviation of degree distribution),
betweenness and clustering coefficient, may reflect the networks
synchronizability to some extent, but which one or ones indicate
the network synchronizability simply and exactly?

There is another problem need to be mentioned. When investigating
the relations between various topological ingredients and network
synchronizability, some parameters, such as rewiring probability
$p$ of WS networks\cite{Betweenness2} or power-law exponent
$\gamma$ of the degree distribution in scale-free
networks\cite{Betweenness1}, are adjusted to modulate other
topological ingredients, like average distance or heterogeneity of
degree distribution. However, in this process all the topological
ingredients keep changing with the adjusting of these parameters.
It is impossible to get clear relation between an ingredient and
synchronizability when other ingredients are still varying,
especially when we do not know the tracks of their motions.

Here in this paper, we try to discuss the relationship between these
ingredients and the synchronization of complex networks precisely.
This paper is organized as follows. In section 2, we give a brief
review on how to measure the network synchronizability. And then, in
section 3, the so-called random interchanging
algorithm\cite{Reshuffle1,Reshuffle2,Reshuffle3} is introduced,
which allows one to manipulate the clustering coefficient while
keeping the network's degree distribution unchanged. The main
simulations are shown in section 4, and an interesting case is laid
out and discussed in section 5. Finally, in section 6, we sum up
this paper and discuss the relevance of our work to the real world.

\begin{figure}
\begin{center}
 \scalebox{0.5}[0.5]{\includegraphics{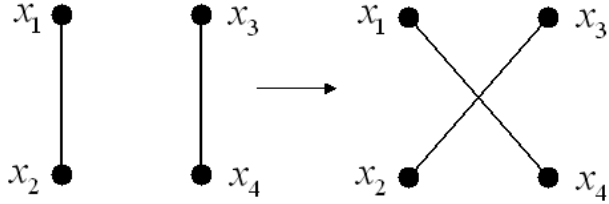}}
\caption{The sketch maps of random interchanging algorithm.}
\end{center}
\end{figure}

\section{Network Synchronizability}
In this section, we will introduce a generic model of coupled
oscillators on networks and a useful measure\cite{master1}, which
is often used to test the stability of the global synchronized
states. Consider $N$ identical dynamical systems (oscillators)
with the same output function, which are located on the vertices
of a network and coupled linearly and symmetrically with
neighbors. The coupling fashion ensures the synchronization
manifold an invariant manifold, and the dynamics can be locally
linearized near the synchronous state. The state of the $i$th
oscillator is denoted by $\textbf{x}^i$, we get the set of
equations of motion governing the dynamics of the $N$ coupled
oscillators:
\begin{equation}
\dot{\textbf{x}}^i=\textbf{F}(\textbf{x}^i)+\eta\sum_{j=1}^NG_{ij}\textbf{H}(\textbf{x}^j),
\end{equation}
where $\dot{\textbf{x}}^i=\textbf{F}(\textbf{x}^i)$ governs the
dynamics of individual oscillator, $\textbf{H}(\textbf{x}^j)$ is
the output function and $\eta$ the coupling strength. The $N\times
N$ Laplacian $\textbf{G}$ is given by
\begin{equation}
    G_{ij}=\left\{
    \begin{array}{cc}
    k_i   &\mbox{for $i=j$}\\
     -1    &\mbox{for $j\in\Lambda_i$}   ,\\
     0    &\mbox{otherwise}
    \end{array}
    \right.
\end{equation}
where $\Lambda_i$ denotes the neighbor set of node $i$. Because of
the positive semidefinite of $\textbf{G}$, all the eigenvalues of
it are nonnegative reals and the smallest eigenvalue $\theta_0$ is
always zero, for the rows of $\textbf{G}$ have zero sum. If all
the nodes are connected, there is only one zero eigenvalue. Thus,
the eigenvalues can be ranked as
$\theta_0<\theta_1\leq\cdots\leq\theta_{N-1}$. The ratio of the
maximum eigenvalue $\theta_{N-1}$ to the smallest nonzero one
$\theta_1$ is widely used to measure the synchronizability of the
network\cite{master1}, if the eigenratio $R=\theta_{N-1}/\theta_1$
satisfies
\begin{equation}
R<\alpha_2/\alpha_1,
\end{equation}
we say the network is synchronizable. The right-hand side
$\alpha_2/\alpha_1$ of this inequality depends on the dynamics of
individual oscillator and the output function (one can see ref.
\cite{master2} for details), while the eigenratio $R$ depends only
on the Laplacian $\textbf{G}$. $R$ indicates the synchronizability
of the network, the smaller it is the better synchronizability and
vice versa. In this paper, for universality, we will not address a
particular dynamical system, but concentrate on how the network
topology affects the eigenratio $R$.

\section{The Random Interchanging Algorithm}
To investigate the structural effects on network
synchronizability, we use random interchanging algorithm
\cite{Reshuffle1,Reshuffle2,Reshuffle3} to adjust clustering
coefficient while keeping degree distribution unchanged. The
procedure is as follows:

(1) Randomly pick two existing edges $e_1=x_1x_2$ and
$e_2=x_3x_4$, such that $x_1\neq x_2\neq x_3\neq x_4$ and there is
no edge between $x_1$ and $x_4$ as well as $x_2$ and $x_3$.

(2) Interchange these two edges, that is, connect $x_1$ and $x_4$
as well as $x_2$ and $x_3$, and remove the edges $e_1$ and $e_2$.

(3) Ensure the network is still connected and compute whether this
interchange increases/decreases the network clustering
coefficient. If it does, accept the new configuration, else
recover the old one.

(4) Repeat step (1) unless the desired clustering coefficient is
achieved.

Since this algorithm only rewires connections and does not change
the degree of any node, the degree distribution as well as the
degree sequence is fixed. Figure 1 provides a sketch maps of
random interchanging algorithm, which may help us understanding
the program flow.

\section{Simulations}
In the random interchanging process, operations that bring
nonlocal couplings will reduce network average distance $L$
\cite{Easier3,HuJK}, and at the same time the clustering
coefficient $C$ will be reduced\cite{Cluster2}. Figure 2 and 3
exhibit the relationship between $L$ and $C$. In figure 2, the
original networks are the WS networks with size $N=2000$, average
degree $<k>=4$, and standard deviations of degree distributions
$\sigma$=0.2, 0.5, 0.6 and 0.87, respectively. In figure 3, the
original networks are the  extensional BA
networks\cite{ExtBA1,ExtBA2} with $N=2000$, average degree
$<k>=12$ and standard deviations of degree distributions
$\sigma$=18.43, 19.26, 20.65 and 21.26, respectively. Here, the
different standard deviations for WS networks and extensional BA
networks are obtained by adjusting the rewiring probability $p$
and the power-law exponent $\gamma$, respectively. Clearly, the
trends of $L$ and $C$ are qualitatively the same. We have checked
that the positive correlation between $L$ and $C$ is not sensitive
to the network size, the average degree and the standard deviation
of degree distribution. In this paper, we examine the relation
between average distance and synchronizability, and the relation
between clustering coefficient and synchronizability can be
obtained easily.

\begin{figure}
\begin{center}
\scalebox{1.2}[1.2]{\includegraphics{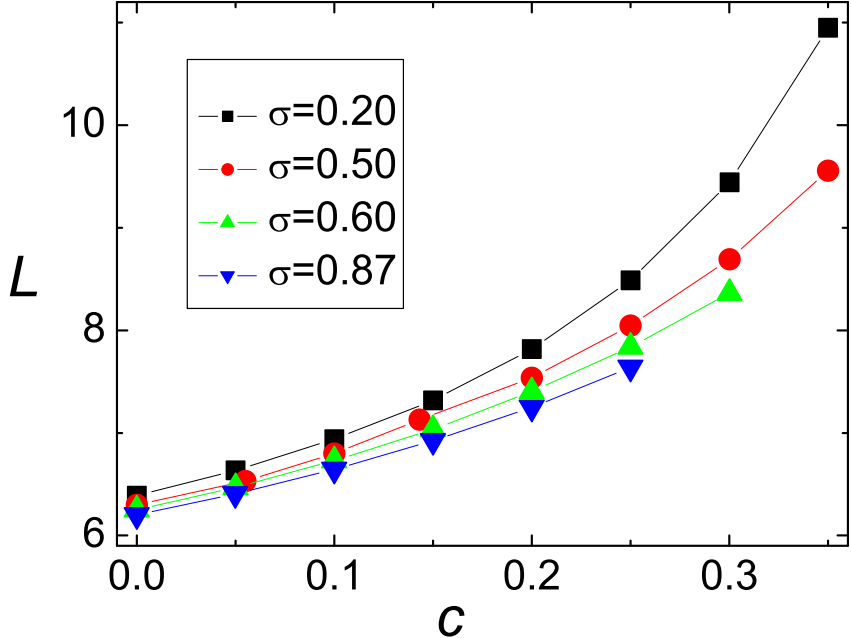}} \caption{(Color
online) The relationship between average distance $L$ and clustering
coefficient $C$ when the original networks are the WS networks. The
black squares, red circles, green up-triangles and blue
down-triangles represent the cases of $\sigma=$0.20, 0.50, 0.60 and
0.87, respectively. All the data are the average over 20 different
realizations.}
\end{center}
\end{figure}

The eigenratios for WS and extensional BA model are obtained
numerical and their behaviors with the average distance $L$ at
different standard deviations of degree distributions are
exhibited in Figs. 4 and 5 respectively. From each curve in Fig. 4
it can be seen that with the increasing of average distance $L$,
the eigenratio $R$ grows, which means shorter average distance
predicts better synchronizability. The similar result is obtained
for extensional BA network shown in Fig. 5. The present result is
consistent with the very recent result\cite{Cluster,WuXiang} that
the larger clustering coefficient will inhibit global
synchronization in scale-free network.

It can also be seen from Figs. 4 and 5 , at equal average distance
$L$ the larger the standard deviation of degree $\sigma$ is, the
larger the eigenratio $R$ will be, indicating networks with a
homogeneous distribution of connectivity are more synchronizable
than heterogeneous ones when average distance $L$ keeps constant.

\begin{figure}
\begin{center}
\scalebox{1.2}[1.2]{\includegraphics{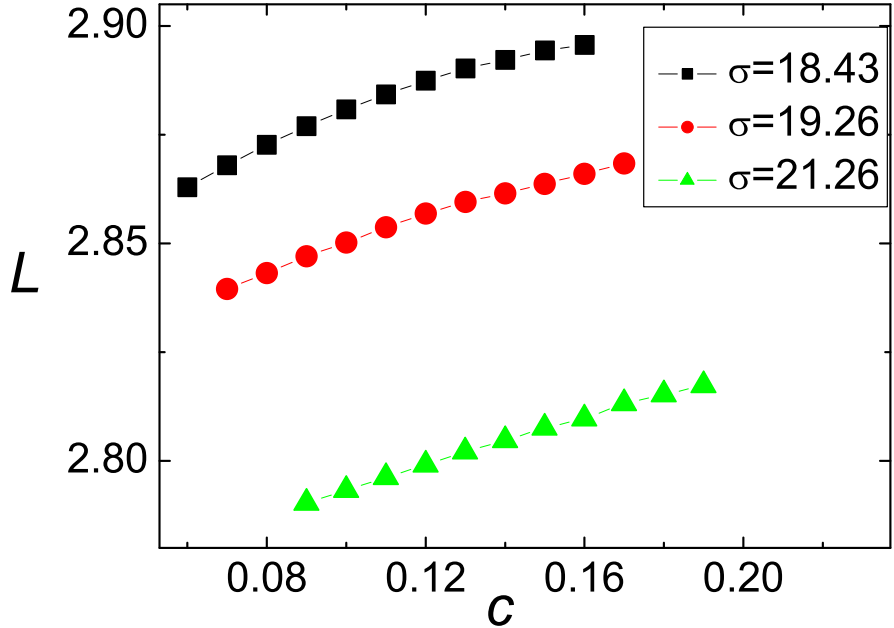}} \caption{(Color
online) The relationship between average distance $L$ and clustering
coefficient $C$ when the original networks are the extensional BA
networks. The black squares, red circles, green up-triangles and
blue down-triangles represent the cases of $\sigma=$18.43, 19.26 and
21.26, respectively. All the data are the average over 20 different
realizations.}
\end{center}
\end{figure}

Average distance and heterogeneity of connections are topological
ingredients for network synchronization. Shortening average
distance and making the connections more homogeneous solely will
increase network synchronizability, however, only their
combination could make the network have strong synchronizability.
The average distance of star coupling network is very short,
$L\rightarrow2$ as $N\rightarrow\infty$, while the standard
deviation of degree is very large, $\sigma\sim\sqrt{N}$, and the
eigenration $R\rightarrow N$, suggesting that it is hard to
synchronize when the network size is large. In one-dimensional
ring lattice, all the nodes have equal degree, thus $\sigma=0$.
However, the average distance $L\sim N/4z$ is too large (here $z$
denotes the coordination number\cite{Newman1999}), thus it is also
very hard to synchronize with the increasing network
size\cite{ex1}.

\begin{figure}
\begin{center}
\scalebox{1.2}[1.2]{\includegraphics{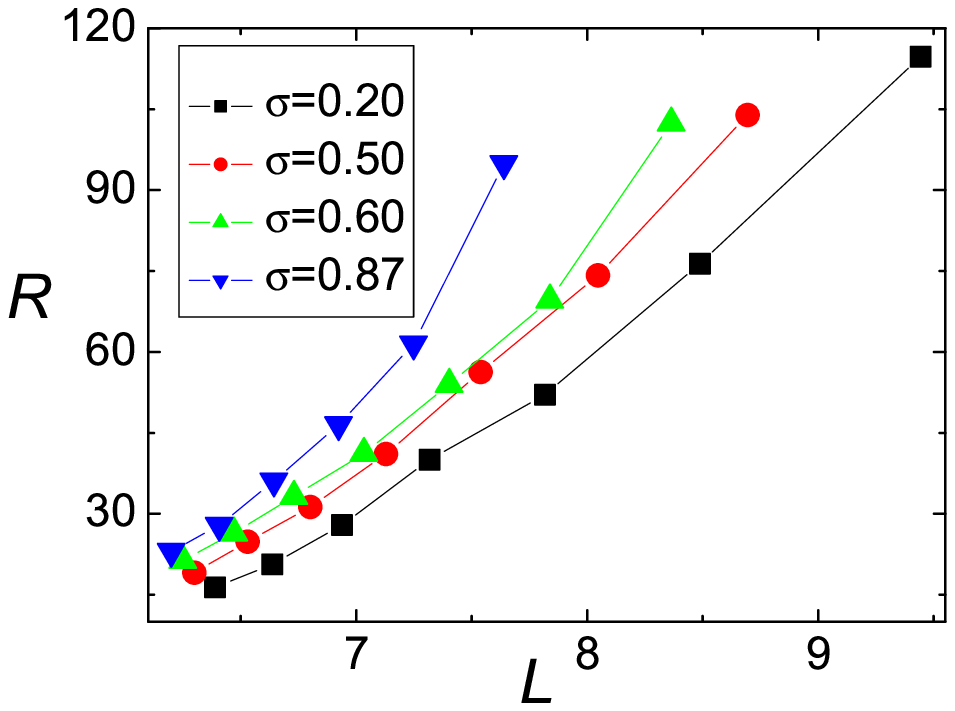}} \caption{(Color
online) The eigenratio $R$ vs average distance $L$ when the original
networks are the WS networks. Eigenratio $R$ shows positive
correlation with average distance $L$ when standard deviation of
degree $\sigma$ is fixed, and at a fixed value of $L$, $R$ will
increase with the rising of $\sigma$. For a variety of chaotic
oscillators, $\alpha_2/\alpha_1$ ranges from 5 to 100\cite{Easier3},
so we only investigate the situations with eigenratios less than
about 100. All the data are the average over 20 different
realizations.}
\end{center}
\end{figure}

For many scale-free network models, heterogeneous distribution of
connectivity tends to reduce the average network
distance\cite{HandL,Betweenness1}, WS small-world network obeys
the same law\cite{Betweenness2}. Thus, in this two kind of
networks, the increasing of heterogeneity will diminish the
average distance. While, for WS small-world network the standard
deviation of degree is very small, the network synchronizability
is mainly determined by average distance, with $L$'s decreasing,
although $\sigma$ increases, $R$ will still be diminished, the
network becomes more synchronizable. For some real-life scale-free
networks, because of their ultra-small feature\cite{HandL,HandZ}
($L\sim \ln\ln N$ or even shorter) and large degree deviation, $R$
strongly depends on heterogeneity, the more heterogeneous is the
harder to synchronize. Therefore, shortening average distance is
an effective way to enhance small-world network synchronizability
and diminishing the heterogeneity of connectivity can make
scale-free network synchronize easier.

\begin{figure}
\begin{center}
\scalebox{1.2}[1.2]{\includegraphics{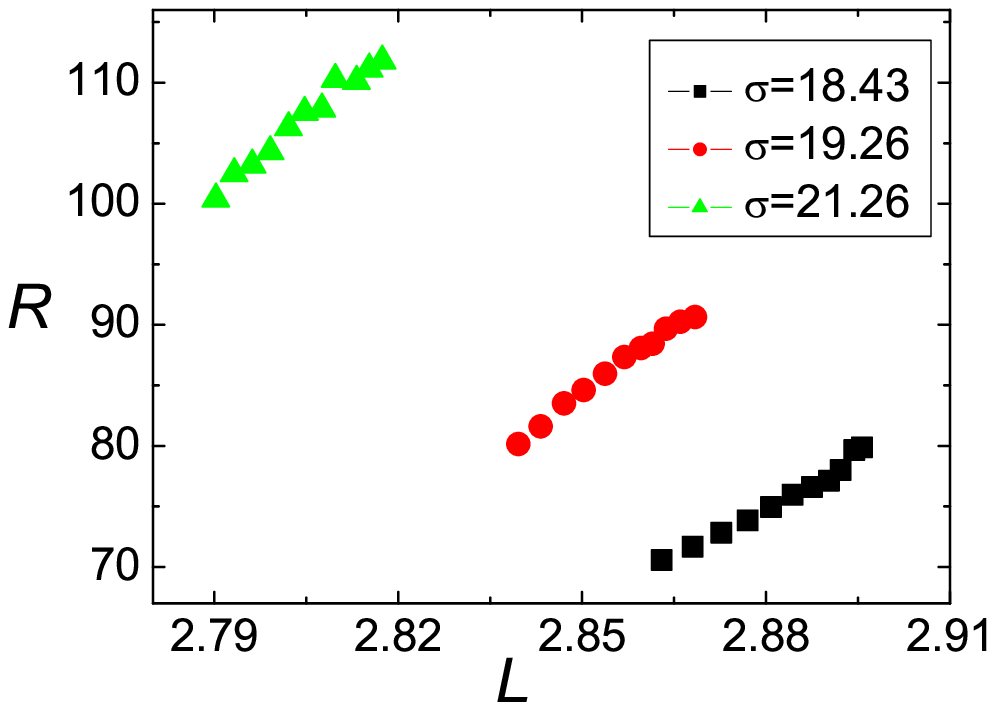}} \caption{(Color
online) The eigenratio $R$ vs average distance $L$ when the original
networks are the extensional BA networks. Eigenratio $R$ shows
positive correlation with average distance $L$ when standard
deviation of degree $\sigma$ is fixed, and at a fixed value of $L$,
$R$ will increase with the rising of $\sigma$. All the data are the
average over 20 different realizations.}
\end{center}
\end{figure}

\section{An Interesting Case: Smaller Maximal Betweenness may not Indicate Better Synchronizability}
Some previous studies suggest that the maximal betweenness
centrality $B_{max}$ is a suitable indicator for predicting
synchronizability on complex
networks\cite{Betweenness2,Cluster,Betweenness3}; the larger
$B_{max}$ is, the poorer the synchronizability. The betweenness
centrality of node $n$ is defined as the probability that a
randomly selected shortest path of a randomly picked pair of nodes
contains the node $n$\cite{BetweennessDef1,BetweennessDef2}
\begin{equation}
B_{n}:=\frac{1}{(N-1)(N-2)}\sum_{i\neq j\neq
n}\frac{g_{ij}(n)}{g_{ij}},
\end{equation}
where $g_{ij}$ is the number of shortest paths between nodes $i$
and $j$, and $g_{ij}(n)$ is the number of those paths passing
through node $n$. From the definition of betweenness centrality,
it is easy to get the relationship between the average betweenness
centrality $<B>$ and the average distance $L$
\begin{equation}
<B>=\frac{N(N-1)(L-1)}{N(N-1)(N-2)}=\frac{L-1}{N-2}.
\end{equation}
Previous studies indicate that there exists strongly positive
correlation between degree and betweenness
centrality\cite{load1,load2}, that is to say, the node with larger
degree will statistically have higher betweenness centrality.
Therefore, betweenness centrality is approximately determined by
average distance and degree heterogeneity: its average value is
determined by the average distance, and its breadth by the
heterogeneity of connectivity. Therefore, $B_{max}$ can reflect
the influences of both the average distance and degree
heterogeneity, which may be the reason why some authors think that
$B_{max}$ is a suitable quantity to estimate network
synchronizability.

\begin{figure}
\begin{center}
\scalebox{1.2}[1.2]{\includegraphics{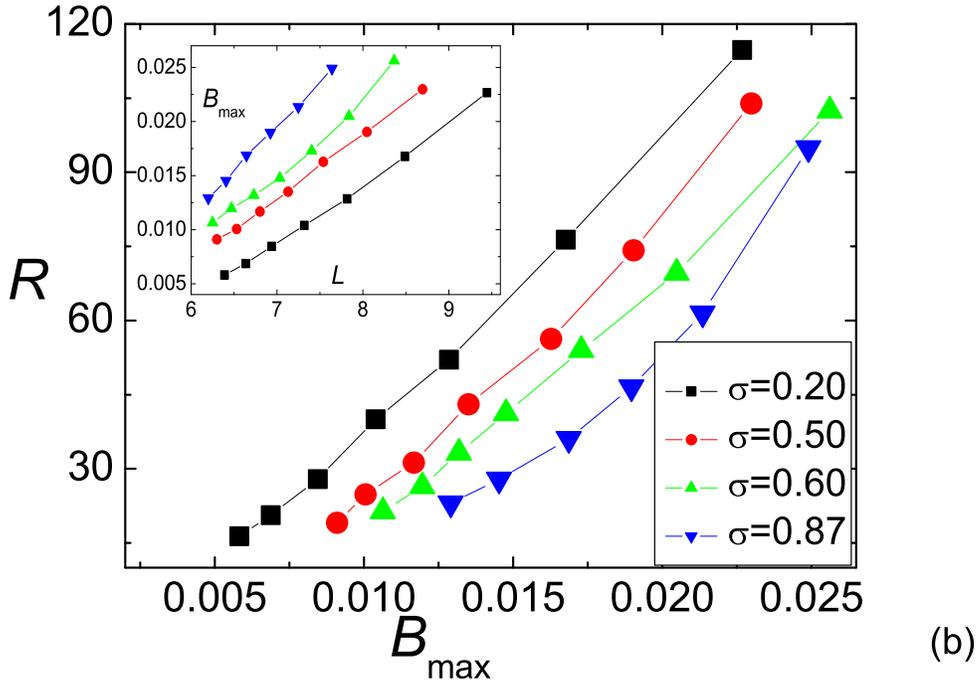}} \caption{(Color
online) The eigenratio $R$ vs maximal betweenness $B_{max}$ when the
original network are the WS networks. The eigenratio $R$ is positive
correlated with maximal betweenness $B_{max}$ when standard
deviation of degree $\sigma$ is fixed. All the data are the average
over 20 different realizations.}
\end{center}
\end{figure}

The insets of figure 6 and 7 respectively show the changes of
maximal betweenness with average distance at different standard
deviation of degree. For WS network, the increasing of average
distance or degree deviation solely always induces the increasing
of maximal betweenness. It is consistent with the former analysis.
However, for extensional BA network, maximal betweenness is not
sensitive to average distance but increases with the increasing of
degree deviation clearly.

\begin{figure}
\begin{center}
\scalebox{1.2}[1.2]{\includegraphics{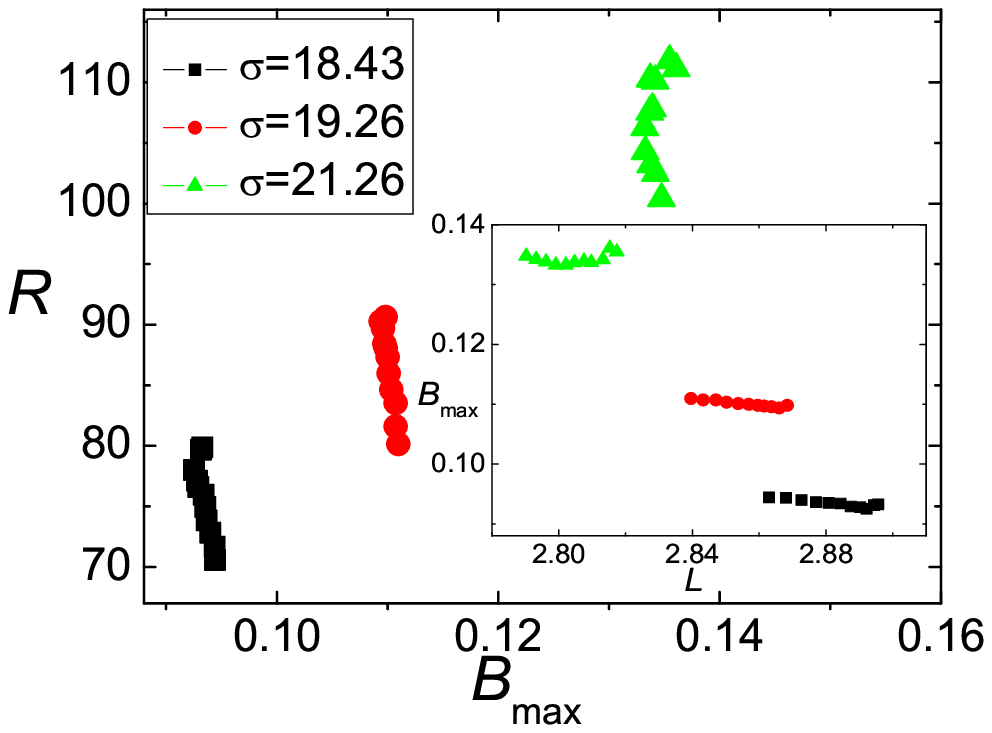}} \caption{(Color
online) The eigenratio $R$ vs maximal betweenness $B_{max}$ when the
original network the extensional BA networks. Clearly, no positive
correlation between $R$ and $B_{max}$ can be observed under this
case. All the data are the average over 20 different realizations.}
\end{center}
\end{figure}

By running the random interchanging algorithm, we computing $R$
and $B_{max}$ for different configurations, figure 6 and 7 show
the correlation between them at fixed $\sigma$ for WS network and
extensional BA network, respectively. When the original network is
a WS network, there exist strongly positive correlation between
$R$ and $B_{max}$ at fixed $\sigma$ (see figure 6), which support
the previous conclusion\cite{Betweenness2}. But for different
degree deviation, networks shows different $R$ at the same
$B_{max}$, although the number of nodes and edges are the same.
When the original network is an extensional BA network at fixed
$\sigma$, the positive correlation between $R$ and $B_{max}$
vanishes. The simulation results show that the single quantity,
maximal betweenness $B_{max}$, may not give a comprehensive
description of network synchronizability.

\section{Conclusion Remarks}
In conclusion, with the help of random interchanging algorithm we
show that shorter average distance and homogeneity solely will
lead to better synchronizability, but only their combination could
make the network easy to synchronize.

Some Numerical studies have been done to check if the maximal
betweenness $B_{max}$ is a proper quantity to estimate network
synchronizability. The simulation results strongly suggest that
the single quantity, $B_{max}$, may not give a comprehensive
description of network synchronizability.

It is worthwhile to emphasize that this work is not only of
theoretical interest, but also of practical value. The clear
picture of topological effects on network synchronizability may
provide us a guideline to design algorithm aiming at enhancing or
reducing the network synchronizability\cite{zz}.

\subsection*{Acknowledgement}
This work is supported by the National Natural Science Foundation of
China under Grant No. 10472116, 70471033, 70571074, 10532060,
10547004 and 70271070, the Specialized Research Fund for the
Doctoral Program of Higher Education (SRFDP No.20020358009), Special
Research Founds for Theoretical Physics Frontier Problems (NSFC
Grant No. A0524701), and Specialized Program under President Funding
of Chinese Academy of Science.


\begin{thebibliography}{1}
\bibitem{ex1} In one-dimensional ring lattice, the eigenratio $R=1/\sin^{2}(\pi z/N)$, thus it is very hard to synchroniza when $N$ gets large.
\end{thebibliography}
\end{document}